\documentstyle[12pt]{article}
\newcommand{\be}{\begin{eqnarray}}
\newcommand{\ee}{\end{eqnarray}}

\def\lsim{\mathrel{\rlap{\lower4pt\hbox{\hskip1pt$\sim$}}
    \raise1pt\hbox{$<$}}}               
\def\gsim{\mathrel{\rlap{\lower4pt\hbox{\hskip1pt$\sim$}}
    \raise1pt\hbox{$>$}}}               

\input{epsfig.sty}

\begin{document}

\Huge{\noindent{Istituto\\Nazionale\\Fisica\\Nucleare}}

\vspace{-3.9cm}

\Large{\rightline{Sezione SANIT\`{A}}}
\normalsize{}
\rightline{Istituto Superiore di Sanit\`{a}}
\rightline{Viale Regina Elena 299}
\rightline{I-00161 Roma, Italy}

\vspace{0.65cm}

\rightline{INFN-ISS 96/8}
\rightline{September 1996}

\vspace{1.25cm}

\begin{center}

\LARGE{Electroproduction of the Roper resonance\\ and the constituent
quark model\footnote{{\bf To appear in Physics Letters B.}}}\\

\vspace{1.0cm}

\large{ F. Cardarelli$^{(1)}$, E. Pace$^{(2)}$, G. Salm\`{e}$^{(3)}$,
S. Simula$^{(3)}$}\\

\vspace{0.25cm}

\normalsize{$^{(1)}$ Dept. of Physics and Supercomputer Computations
Research Institute\\ Florida State University, Tallahassee, FL 32306,
USA\\ $^{(2)}$ Dipartimento di Fisica, Universit\`{a} di Roma "Tor
Vergata" \\ and Istituto Nazionale di Fisica Nucleare, Sezione Tor
Vergata\\ Via della Ricerca Scientifica 1, I-00133 Roma, Italy \\ 
$^{(3)}$ Istituto Nazionale di Fisica Nucleare, Sezione Sanit\`{a}\\
Viale Regina Elena 299, I-00161 Roma, Italy}

\end{center}

\vspace{0.5cm}

\begin{abstract}

\noindent A parameter-free evaluation of the $N - P_{11}(1440)$
electromagnetic transition form factors is performed within a
light-front constituent quark model, using for the first time the
eigenfunctions of a mass operator which generates a large amount of
configuration mixing in baryon wave functions. A one-body
electromagnetic current, which includes the phenomenological
constituent quark form factors already determined from an analysis of
pion and nucleon experimental data, is adopted. For $Q^2$ up to few
$(GeV/c)^2$, at variance with the enhancement found in the elastic
channel, the effect of configuration mixing results in a significant
suppression of the calculated helicity amplitudes with respect to both
relativistic and non-relativistic calculations, based on a simple
gaussian-like ansatz for the baryon wave functions. 

\end{abstract}

\vspace{0.5cm}


\vspace{1cm}

\newpage

\pagestyle{plain}

\indent The investigation of the electromagnetic (e.m.) excitations of
nucleon resonances can shed light on their structure in terms of quarks
and gluons. In this respect, the Roper resonance, $P_{11}(1440)$, plays
a particular role. Within the constituent quark ($CQ$) picture (see,
e.g., \cite{KI80,CK95}) this resonance is commonly assigned to a radial
excitation of the nucleon, whereas it has been argued
\cite{BC83,HYBRID,CM91} that it might be a hybrid state, containing an
explicit excited glue-field configuration (i.e., a $q^3G$ state).
Within the $q^3$ assignment the spin-flavour part of the
Roper-resonance wave function is commonly considered to be identical to
that of the nucleon, whereas the $q^3G$ state is directly orthogonal to
the nucleon in the spin-flavour space. Then, it is expected
\cite{HYBRID} that such different spin structures of the Roper
resonance could lead to different behaviours of its e.m. helicity
amplitudes as a function of the four-momentum transfer, so that future
experiments planned at $TJNAF$ \cite{TJNAF} might provide signatures
for hybrid baryons. However, the predictions of Ref. \cite{HYBRID} have
been obtained within a non-relativistic framework and using simple
gaussian-like wave functions. Within the $CQ$ model, the relevance of
the relativistic effects on the helicity amplitudes of the Roper
resonance has been illustrated by Weber \cite{WE90} and by Capstick and
Keister \cite{CK95}, where (we stress) gaussian-like wave functions
were still adopted. 
  
\indent The purpose of this letter is to compute the e.m. $N -
P_{11}(1440)$ transition form factors in the relativistic $CQ$ model
developed in \cite{CAR_M,CAR_N,CAR_ND}. The model incorporates the
following features: ~ i) a proper treatment of relativistic effects
through the light-front ($LF$) formalism; ~ ii) the use of the
eigenfunctions of a baryon mass operator having a much closer
connection to the mass spectrum with respect to a gaussian-like ansatz;
~ iii) the use of a one-body approximation for the e.m. current able to
reproduce the experimental data on the nucleon form factors. Inside
baryons the $CQ$'s are assumed to interact via the $q - q$ potential of
Capstick and Isgur ($CI$) \cite{CI86}. A relevant feature of this
interaction is the presence of an effective one-gluon-exchange ($OGE$)
term, which produces a huge amount of high-momentum components and
$SU(6)$ breaking terms in the baryon wave functions (see
\cite{CAR_N,CAR_ND}); in what follows we will refer to these effects as
the configuration mixing. Finally, an effective one-body e.m. current,
including Dirac and Pauli form factors for the $CQ$'s (cf. also Ref.
\cite{FH82}), is adopted. The $CQ$ form factors have been determined in
\cite{CAR_N} using as constraints the pion and nucleon experimental
data. In \cite{CAR_ND} our parameter-free prediction for the magnetic
form factor of the $N - \Delta(1232)$ transition has been checked
against available data. In this letter, our parameter-free results for
the $N - P_{11}(1440)$ helicity amplitudes will be presented, showing
that the configuration mixing leads to a significant suppression of the
calculated helicity amplitudes with respect to relativistic as well as
non-relativistic calculations, based on a simple gaussian-like ansatz
for the wave functions. 

\indent In the $LF$ hamiltonian dynamics (cf. \cite{KP91}) intrinsic
momenta of the $CQ$'s, $k_i$, can be obtained from the on-mass-shell
momenta $p_i$ in a general reference frame, through the $LF$ boost
$L_f^{-1}(P_0)$, which transforms the momentum $P_0 \equiv \sum_{i=1}^3
p_i$ as $L_f^{-1}(P_0) ~ P_0 = (M_0, 0, 0, 0)$ without Wigner
rotations. Thus, one has $k_i = L_f^{-1}(P_0) ~ p_i$ and, obviously,
$\sum_{i=1}^3 ~ \vec{k}_i = 0$. In this formalism a baryon state in the
$u  -d$ sector, $|\Psi_{J J_n ~ \pi}^{T T_3}, ~ \tilde{P} \rangle$, is
an eigenstate of: i) isospin, $T$ and $T_3$; ii) parity, $\pi$; iii)
kinematical (non-interacting) $LF$ angular momentum operators $j^2$ and
$j_n$, where the vector $\hat{n} = (0,0,1)$ defines the spin
quantization axis; iv) total $LF$ baryon momentum $\tilde{P} \equiv
(P^+, \vec{P}_{\perp}) = \tilde{p}_1 + \tilde{p}_2 + \tilde{p}_3$,
where $P^+ = P^0 + \hat{n} \cdot \vec{P}$ and $\vec{P}_{\perp} \cdot
\hat{n} = 0$. We explicitly construct $|\Psi_{J J_n ~ \pi}^{T T_3}, ~
\tilde{P} \rangle$ as eigenstate of an intrinsic $LF$ mass operator,
${\cal{M}} = M_0 + {\cal{V}}$, where $M_0 = \sum_{i=1}^3 \sqrt{m_i^2 +
\vec{k}_i^2 }$ is the free mass operator, $m_i$ the $CQ$ mass and
${\cal{V}}$ a  Poincar\'e invariant interaction. The state $|\Psi_{J J_n
~ \pi}^{T T_3}, ~ \tilde{P} \rangle$ factorizes into $|\Psi_{J J_n ~
\pi}^{T T_3} \rangle ~ |\tilde{P} \rangle$ and the intrinsic $LF$
angular momentum eigenstate $|\Psi_{J J_n ~ \pi}^{T T_3} \rangle$ can be
constructed from the eigenstate $|\psi_{J J_n ~ \pi}^{T T_3} \rangle$
of the {\em canonical} angular momentum, i.e. $|\Psi_{J J_n ~ \pi}^{T
T_3} \rangle = {\cal{R}}^{\dag} ~ |\psi_{J J_n ~ \pi}^{T T_3} \rangle$,
by means of the unitary operator ${\cal{R}}^{\dag} = \prod_{j=1}^3
R_{Mel}^{\dag} (\vec{k}_{j}, m_j)$, with $R_{Mel} (\vec{k}_{j}, m_j)$
being the generalized Melosh rotation \cite{KP91}. One gets
 \be
    (M_0 + V) ~ |\psi_{J J_n ~ \pi}^{T T_3} \rangle ~ = ~ M ~ |\psi_{J
    J_n ~ \pi}^{T T_3} \rangle
    \label{2}
 \ee
where $M$ is the baryon mass. The interaction $V = {\cal{R}}
{\cal{VR}}^{\dag}$ has to be independent of the total momentum $P$ and
invariant upon spatial rotations and translations (cf. \cite{KP91}). We
can identify Eq. (\ref{2}) with the baryon mass equation proposed by
Capstick and Isgur in \cite{CI86}. The $CI$ effective interaction $V =
\sum_{i<j} V_{ij}$ is composed by a linear confining term (dominant at
large separations) and a $OGE$ term (dominant at short separations). The
latter contains both a central Coulomb-like potential and a
spin-dependent part, responsible for the hyperfine splitting of baryon
masses. The values $m_u = m_d = 0.220 ~ GeV$ \cite{CI86} have been
adopted throughout this work. As in Refs. \cite{CAR_N,CAR_ND}, the mass
equation (\ref{2}) has been solved by expanding the state $|\psi_{J J_n
~ \pi}^{T T_3} \rangle$ onto a (truncated) set of harmonic oscillator
($HO$) basis states and by applying the Rayleigh-Ritz variational
principle. We have included in the expansion all the $HO$ basis states
up to $20 ~ HO$ quanta and the obtained eigenvalues are in agreement
with the results of Ref. \cite{CI86}. The $S$, $S'$ and $D$ components
have been considered and the corresponding probabilities are: $P_S^N =
98.1 \%$, $P_{S'}^N = 1.7 \%$, $P_D^N = 0.2 \%$ for the nucleon and
$P_S^{Roper} = 90.6 \%$, $P_{S'}^{Roper} = 9.3 \%$, $P_D^{Roper} = 0.1
\%$ for the Roper resonance. Note that in \cite{IK79} an approximate
treatment of the hyperfine $OGE$ term led to: $P_S^{Roper} \simeq 97
\%$, $P_{S'}^{Roper} \simeq 3 \%$, $P_D^{Roper} \simeq 0.01 \%$.
Finally, $P$ partial waves have been neglected, because they do not
couple to the main components of the wave functions.  

\indent Let us now consider the $CQ$ momentum distribution $n(p)$,
defined as in \cite{CAR_ND}. The momentum distribution $n(p)$, times
$p^2$, obtained for the nucleon and the Roper resonance using the $CI$
interaction, is shown in Fig. 1(a) and compared with the gaussian-like
ansatz adopted in \cite{CK95}. It can clearly be seen that the
high-momentum tail of both baryon wave functions is sharply enhanced by
the effects due to the $OGE$ interaction. The contributions of the $S$,
$S'$ and $D$ partial waves to the $CQ$ momentum distribution are
separately shown in Fig. 1(b). It turns out that in case of the Roper
resonance the mixed-symmetry $S'$-wave, which has a spin-flavour
structure orthogonal to that of the symmetric $S$-wave component,
yields a significant contribution in a wide range of momenta; moreover,
for $p \lsim 1 ~ GeV/c$ the $S'$ component is much larger in the Roper
resonance than in the nucleon. On the contrary, the $D$-wave components
of both the nucleon and the Roper resonance give a negligible
contribution to $n(p)$. Therefore, in the calculation of the $N -
P_{11}(1440)$ transition form factors we will neglect the contribution
of $D$-wave components. The results reported in  Fig. 1 clearly show
that, when the $OGE$ interaction is fully considered, the resulting
$CQ$ structure of the Roper resonance contains high radial excitations
and sizable mixed-symmetry components, so that it can hardly be
interpreted as a simple (first) radial excitation of the nucleon.
 
\indent The matrix elements of the e.m. $N - P_{11}(1440)$ transition
current can be written as follows (cf., e.g., \cite{WE90})
 \be
    & & \langle \Psi_{{1 \over 2} \nu^* ~ +1}^{{1 \over 2} \tau^*},
    \tilde{P}^* | ~ {\cal{I}}^{\mu}(0) ~ |\Psi_{{1 \over 2} \nu ~
    +1}^{{1 \over 2} \tau}, \tilde{P} \rangle = \delta_{\tau^* \tau} ~
    {\cal{I}}^{\mu}_{\nu^* \nu}(\tau) = \nonumber \\
    & & \delta_{\tau^* \tau} ~ \bar{u}(\tilde{P}^*, \nu^*) \left \{
    F_1^{* \tau}(Q^2) ~ [\gamma^{\mu} + q^{\mu} {M^* - M \over Q^2}] +
    F_2^{* \tau}(Q^2) ~ {i \sigma^{\mu \rho} q_{\rho} \over M^* + M}
    \right \} u(\tilde{P}, \nu)
    \label{3}
 \ee
where $Q^2 \equiv - q \cdot q$ is the squared four-momentum transfer,
$\sigma^{\mu \rho} = {i \over 2}[\gamma^{\mu},\gamma^{\rho}]$,
$u(\tilde{P}, \nu)$ [$u(\tilde{P}^*, \nu^*)$] the nucleon
[Roper-resonance] spinor, $F_{1(2)}^{* \tau}(Q^2)$ the Dirac (Pauli)
form factor associated to the $N - P_{11}(1440)$ transition and $\tau =
\mp 1/2$ (or $\tau = n,p$). In Eq. (\ref{3}) the structure
$\gamma^{\nu} + q^{\mu} (M^* - M) / Q^2$ is required in order to keep
gauge invariance. In the $LF$ formalism (cf. \cite{KP91}) the
space-like e.m. form factors are related to the matrix elements of the
{\em plus} component of the e.m. current (${\cal{I}}^+$) and, moreover,
the choice $q^+ = P^{*+} - P^+ = 0$ allows to suppress the contribution
of the pair creation from the vacuum \cite{ZGRAPH}. The matrix elements
${\cal{I}}^+_{\nu^* \nu}(\tau)$ can be cast in the form
${\cal{I}}^+_{\nu^* \nu}(\tau) =$ $F_1^{* \tau}(Q^2) \delta_{\nu^* \nu}
-$ $F_2^{* \tau}(Q^2) ~ i(\sigma_2)_{\nu^* \nu}$ $Q / (M^* + M)$, where
$\sigma_2$ is a Pauli matrix. Then, the transition form factors
$F_{1(2)}^{* \tau}(Q^2)$ are given by $F_1^{* \tau}(Q^2) =$
$Tr[{\cal{I}}^+(\tau)] / 2$ and $F_2^{* \tau}(Q^2) =$ $i (M^* + M) ~
Tr[\sigma_2 ~ {\cal{I}}^+(\tau)] / 2Q$.

\indent The $N - P_{11}(1440)$ transition form factors will be evaluated
using the eigenvectors of Eq. (\ref{2}) and the plus component of the
one-body e.m. current of Ref. \cite{CAR_N}, viz. 
 \be
    {\cal{I}}^+(0) = \sum_{j=1}^3 ~ I^+_{j}(0) = \sum_{j=1}^3 ~ \left (
    e_j \gamma^+ f_1^j(Q^2) ~ + ~ i \kappa_j {\sigma^{+ \rho} q_{\rho}
    \over 2 m_j}f_2^j(Q^2) \right )
    \label{5}
 \ee
where $e_j$ ($\kappa_j$) is the charge (anomalous magnetic moment) of
the j-th quark, and $f_{1(2)}^j$ its  Dirac (Pauli) form factor. Though
the full hadron e.m. current has to include two-body components for
fulfilling gauge and rotational invariances (see \cite{KP91}), we have
shown \cite{CAR_N} that the effective one-body current component
(\ref{5}) is able to give a coherent description of both the pion and
nucleon experimental form factors. Moreover, using the $CQ$ form factors
determined in \cite{CAR_N}, our parameter-free prediction for the
magnetic form factor of the $N - \Delta(1232)$ transition has been
checked against available data (see \cite{CAR_ND}). Let us stress
that, since our one-body approximation refers to the ${\cal{I}}^+$
component of the current only, with a suitable definition of
the other components the e.m. current can fulfil gauge invariance.

\indent Our results for the magnetic transition form factor
$G_M^{*p}(Q^2) \equiv$ $F_1^{*p}(Q^2) +$ $F_2^{*p}(Q^2)$, obtained using
the $CI$ wave functions both with and without the $CQ$ form factors of
Refs. \cite{CAR_N,CAR_ND}, are shown in Fig. 2 for $Q^2$ up to few
$(GeV/c)^2$ (i.e., in a range of values of $Q^2$ of interest to
$TJNAF$) and compared with the predictions of the relativistic $q^3$
model of Ref. \cite{CK95}, where a gaussian-like ansatz is adopted for
the baryon wave functions and point-like $CQ$'s are assumed. As in the
case of the elastic $G_M^{p}(Q^2)$ form factor (cf. Ref. \cite{CAR_N}),
$G_M^{*p}(Q^2)$ is remarkably sensitive to configuration mixing effects.
However, for $Q^2$ up to few $(GeV/c)^2$ it turns out that the
configuration mixing does not produce in $G_M^{*p}(Q^2)$ the large
enhancement found in the elastic channel. Then, when the $CQ$ form
factor of Ref. \cite{CAR_N} are included, our full prediction and the
one of Ref. \cite{CK95} turn out to be quite similar for the proton,
but strongly different for the $p - P_{11}(1440)$ transition. In
particular, for $Q^2 \sim 1 \div 4 ~ (GeV/c)^2$ our magnetic form
factor $G_M^{*p}(Q^2)$ is suppressed with respect to the prediction of
Ref. \cite{CK95} by a large factor ($\sim 3 \div 4$), which implies a
reduction of about one order of magnitude for the electroproduction
cross section of the Roper resonance.

\indent In what follows, our results will be shown in terms of the
helicity amplitudes $A_{1 \over 2}^{\tau}(Q^2)$ and $S_{1 \over
2}^{\tau}(Q^2)$, defined as 
 \be
    A_{1 \over 2}^{\tau}(Q^2) = {\cal{N}}(Q^2) ~ G_M^{* \tau}(Q^2)
    ~~~~, ~~~~
    S_{1 \over 2}^{\tau}(Q^2) & = & {\cal{N}}(Q^2) {\sqrt{2 K^- K^+}
    \over Q^2} {M^* + M \over 4M^*} ~ G_E^{* \tau}(Q^2)
    \label{6} 
 \ee
where ${\cal{N}} \equiv \sqrt{{\pi \alpha \over K^*} ~ {K^- \over M^*
M}}$, $K^{\pm} \equiv Q^2 + (M^* \pm M)^2$, $K^* \equiv (M^{*2} -
M^2) / 2M^*$ and $G_E^{* \tau} \equiv F_1^{* \tau} - Q^2 ~ F_2^{*
\tau} / (M^* + M)^2$. Our parameter-free predictions for $A_{1 \over
2}^{p(n)}(Q^2)$ and $-S_{1 \over 2}^{p(n)}(Q^2)$ are shown in Fig. 3
and compared with the photoproduction values \cite{PDG} and the results
of phenomenological analyses \cite{GE80,BK86} of available
electroproduction data, as well as with the predictions of the
relativistic $q^3$ model of Ref. \cite{CK95} and of the
non-relativistic $q^3$ and $q^3G$ models of Ref. \cite{HYBRID}(c).
Moreover, in order to better illustrate the effects of the
configuration mixing, the result obtained excluding the $S'$ component
of the $CI$ wave functions of both the nucleon and the Roper resonance,
is also reported in Fig. 3. As in case of $G_M^{*p}(Q^2)$, our results
both for the transverse $A_{1 \over 2}^{p(n)}(Q^2)$ and the longitudinal
$S_{1 \over 2}^p(Q^2)$ helicity amplitudes exhibit a remarkable
reduction with respect to non-relativistic as well as relativistic
predictions, based on simple gaussian-like wave functions. Such a
reduction brings our predictions closer to the results of the
phenomenological analyses of Refs. \cite{GE80,BK86}\footnote{It should
be reminded that model-dependent assumptions made in Refs.
\cite{GE80,BK86} might produce significant uncertainties in the data
analyses.}. At the photon point it can be seen that: ~ i) our
prediction for $A_{1 \over 2}^n(Q^2 = 0)$ agrees well with the $PDG$
value \cite{PDG}, while the absolute value of $A_{1 \over 2}^p(Q^2 =
0)$ is underestimated; ~ ii) the longitudinal helicity amplitudes $S_{1
\over 2}^p(Q^2 = 0)$ and $S_{1 \over 2}^n(Q^2 = 0)$ are remarkably
sensitive to the presence of the mixed-symmetry $S'$ component in the
$CI$ wave functions. The latter feature holds as well up to $Q^2 \sim$
few $(GeV/c)^2$, whereas the transverse helicity amplitudes $A_{1 \over
2}^{p(n)}(Q^2)$ are only slightly modified by the $S'$ partial waves.
Finally, it turns out that the relativistic predictions of the ratio
$A_{1 \over 2}^n(Q^2) / A_{1 \over 2}^p(Q^2)$ differ remarkably from the
non-relativistic result of the $q^3$ and $q^3G$ models (i.e., $A_{1
\over 2}^n(Q^2) / A_{1 \over 2}^p(Q^2) = - 2 / 3$). This result, which
is clearly crucial in a comparison with experimental data, is mainly
due to $S'$ components and to kinematical relativistic effects
associated to the Melosh rotations; in particular, at the photon point
we obtain: $A_{1 \over 2}^n / A_{1 \over 2}^p \simeq - 4/3$ and $\simeq
-1.1$ with and without the $S'$ components, respectively.

\indent Recently \cite{HYBRID}(c), it has been argued that the
uncertainties related to the lack of a precise knowledge of the baryon
wave functions might cancel out in the ratio between transverse and
longitudinal helicity amplitudes. Thus, in order to check this
point, our predictions for the ratio $A_{1 \over 2}^p(Q^2) / [- S_{1
\over 2}^p(Q^2)]$ are shown in Fig. 4 and compared with the results of
non-relativistic \cite{HYBRID}(c) and relativistic \cite{CK95}
calculations, based on a simple gaussian-like ansatz for the baryon
wave functions. It can clearly be seen that up to $Q^2 \sim 2 ~
(GeV/c)^2$ the ratio exhibits a small sensitivity to configuration
mixing effects as well as to the e.m. structure of the $CQ$'s, whereas
it is strongly modified by relativistic effects. In this respect we
want to stress that the relevance of the effects due to the
relativistic compositions of the $CQ$ spins, firstly shown in
\cite{WE90} and clearly exhibited in Figs. 3 and 4, suggests that these
effects, as well as those arising from the configuration mixing, should
be fully included in the predictions of the hybrid $q^3G$ model, before
a meaningful comparison with our light-front $CQ$ picture can be
performed. Finally, note that for $Q^2 \sim 0.2 \div 0.6 ~ (GeV/c)^2$
the relativistic predictions of the transverse amplitudes change sign,
independently of the effects from the configuration mixing and the $CQ$
form factors; therefore, for $Q^2 \sim 0.2 \div 0.6 ~ (GeV/c)^2$ the
Roper-resonance production cross section is expected to be mainly
governed by its longitudinal helicity amplitude.  

\indent In conclusion, the $N - P_{11}(1440)$ electromagnetic transition
form factors have been analyzed within a light-front constituent quark
model, using for the first time baryon wave functions, which incorporate
the configuration mixing generated by the effective one-gluon-exchange
potential of Ref. \cite{CI86}, and a one-body electromagnetic current,
which includes the phenomenological constituent quark form factors
determined in \cite{CAR_N} from an analysis of pion and nucleon
experimental data. It has been shown that the effects of the
configuration mixing (i.e., high-momentum components and $SU(6)$
breaking terms) in the Roper-resonance wave function are large and
prevent to consider the structure of this resonance as a simple (first)
radial excitation of the nucleon. It has been found that the
configuration mixing yields a remarkable suppression of the calculated
helicity amplitudes with respect to relativistic and non-relativistic
predictions, based on a simple gaussian-like ansatz for the baryon wave
functions. Moreover, the longitudinal helicity amplitudes exhibit an
appreciable sensitivity to the mixed-symmetry $S'$ components,
generated in the baryon wave functions by the hyperfine interaction.

\vspace{0.25cm}

{\bf Acknowledgments.} One of the authors (F.C.) acknowledges the
partial support by the U.S. DOE through Contract DE-FG05-86ER40273, and
by the SCRI of the Florida State University, partially funded through
the Contract DE-FC05-85ER250000.

\newpage

\begin{center}

{\bf Figure Captions}

\end{center}

\vspace{0.5cm}

Fig. 1. (a) The momentum distribution $n(p)$ of the constituent quarks
in the nucleon (thick lines) and in the Roper resonance (thin lines),
times $p^2$. The solid and dashed lines are the $CQ$ momentum
distributions obtained from the eigenstates of Eq. (\ref{2}) with the
$CI$ interaction \cite{CI86} and those corresponding to the
gaussian-like ansatz, adopted in \cite{CK95}, respectively. ~ (b)
Contributions of various partial waves to the $CQ$ momentum
distribution (times $p^2$) in the nucleon (thick lines) and in the
Roper resonance (thin lines), obtained using the $CI$ interaction. The
solid, dashed and dot-dashed lines correspond to the $S$, $S'$ and $D$
partial-wave contributions, respectively.

\vspace{0.5cm}

Fig. 2. The magnetic form factor $G_M^{*p}(Q^2)$ for the $p -
P_{11}(1440)$ transition versus $Q^2$. The solid line is our
prediction, obtained using the eigenstates of the mass equation
(\ref{2}) with the $CI$ interaction and the one-body current component
(\ref{5}) with the $CQ$ form factors of Ref. \cite{CAR_N}. The dotted
line is obtained with the $CI$ wave functions, but assuming point-like
$CQ$'s (i.e., putting in Eq. (\ref{5}) $f_1^j = 1$ and $\kappa_j = 0$).
The dot-dashed line is the result of Ref. \cite{CK95}, obtained using a
simple gaussian-like ansatz for the baryon wave functions and assuming
point-like $CQ$'s.

\vspace{0.5cm}

Fig. 3. The $N - P_{11}(1440)$ helicity amplitudes $A_{1 \over
2}^{p(n)}(Q^2)$ and $-S_{1 \over 2}^{p(n)}(Q^2)$, as a function of
$Q^2$. The full dots are the $PDG$ values \cite{PDG}, while the full
squares and open dots are the results of the analysis of available
electroproduction data performed in Refs. \cite{GE80} and \cite{BK86},
respectively. Thick lines correspond to the results of $LF$
calculations. The solid and dot-dashed lines are the same as in Fig.
2. The dashed lines are the results of our calculations performed
excluding the $S'$-wave components of the $CI$ wave functions of both
the nucleon and the Roper resonance. Thin lines are the results of
non-relativistic calculations of Ref. \cite{HYBRID}(c). The
long-dashed and dot-dashed lines correspond to the $q^3G$ and $q^3$
models, evaluated using Eqs. (5) and (8) of Ref. \cite{HYBRID}(c),
respectively. Note that within the hybrid $q^3G$ model $S_{1 \over
2}^{p(n)}(Q^2) = 0$, whereas only $S_{1 \over 2}^n(Q^2)$ is vanishing
within the non-relativistic $q^3$ model. In (b) and (d) the error bars
on the solid thick line represent the uncertainties related to the
numerical Monte Carlo integration procedure.

\vspace{0.5cm}

Fig. 4. Ratio of the transverse $A_{1 \over 2}^p(Q^2)$ to the
longitudinal $-S_{1 \over 2}^p(Q^2)$ helicity amplitudes of the $p -
P_{11}(1440)$ transition, as a function of $Q^2$. Thick lines correspond
to the results of $LF$ calculations. The solid, dashed, dotted and
dot-dashed lines are the same as in Figs. 2 and 3. The thin dot-dashed
line is the prediction of the non-relativistic $q^3$ model of Ref.
\cite{HYBRID}(c). The erros bars are as in Fig. 3.

\newpage

\begin{figure}

\vspace{-4.0cm}

\epsfxsize=11.5cm \epsfig{file=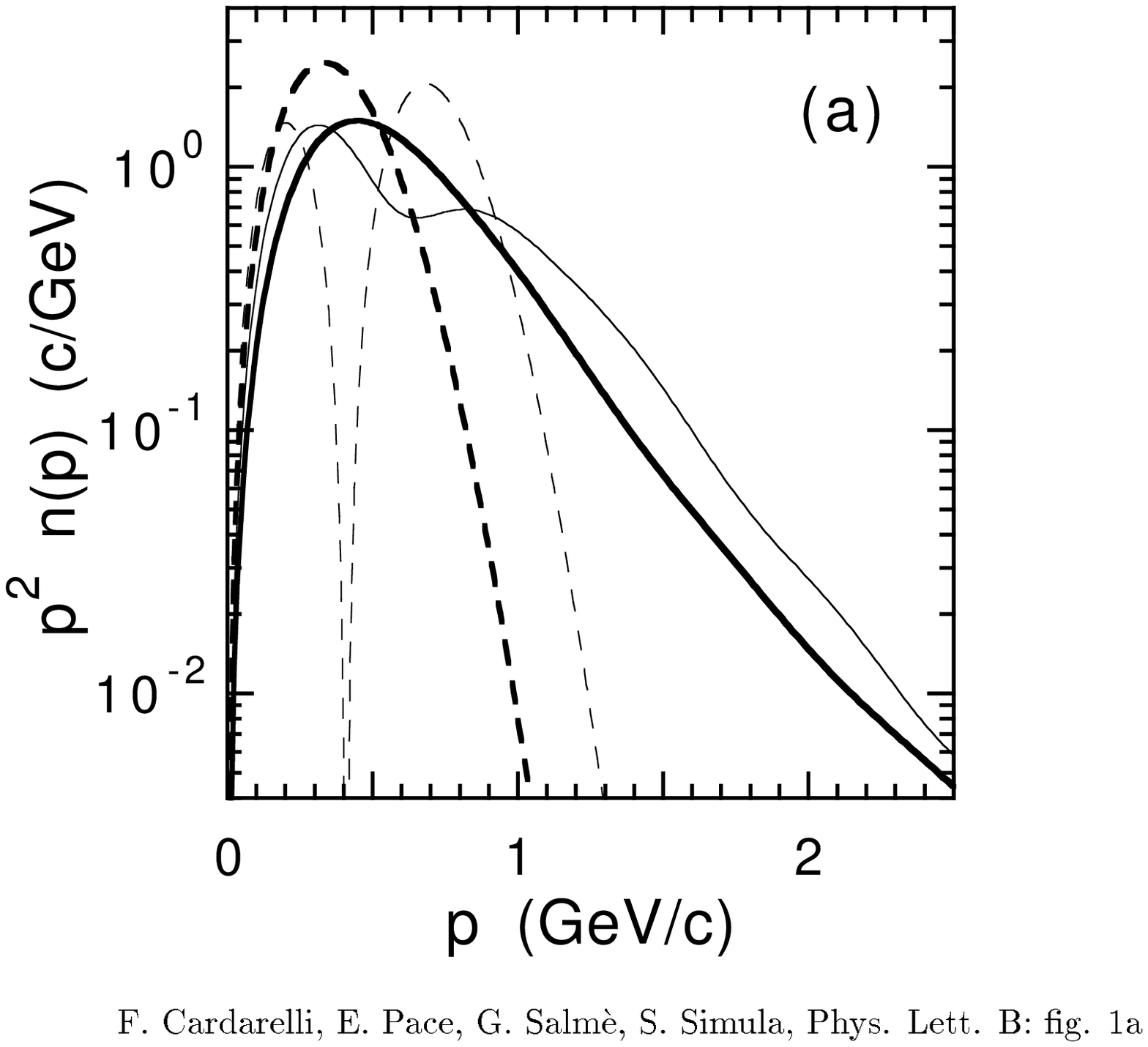}

\end{figure}

\begin{figure}

\vspace{-2.5cm}

\epsfxsize=11.5cm \epsfig{file=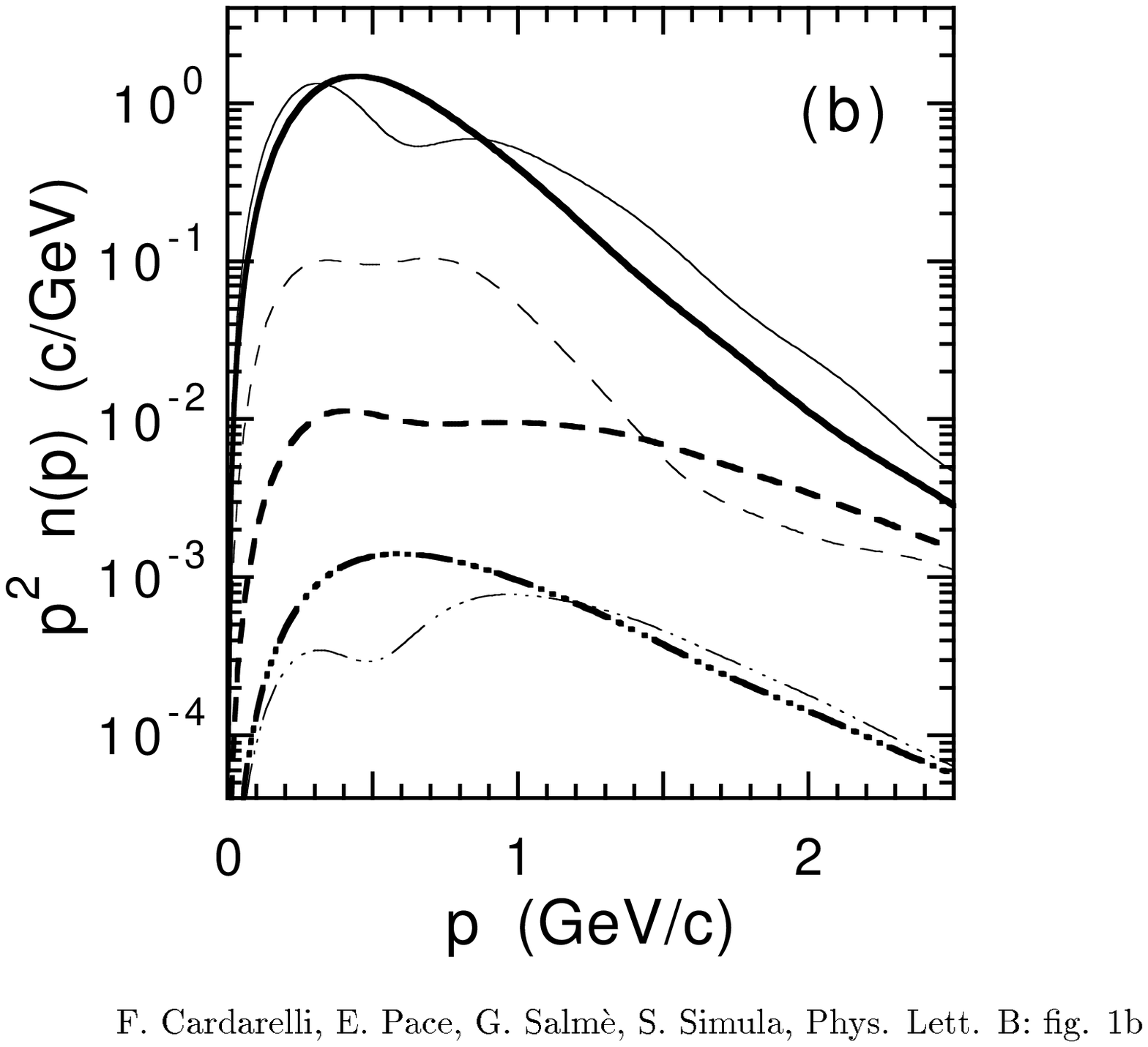}

\end{figure}

\newpage

\begin{figure}

\vspace{-2.5cm}

\epsfig{file=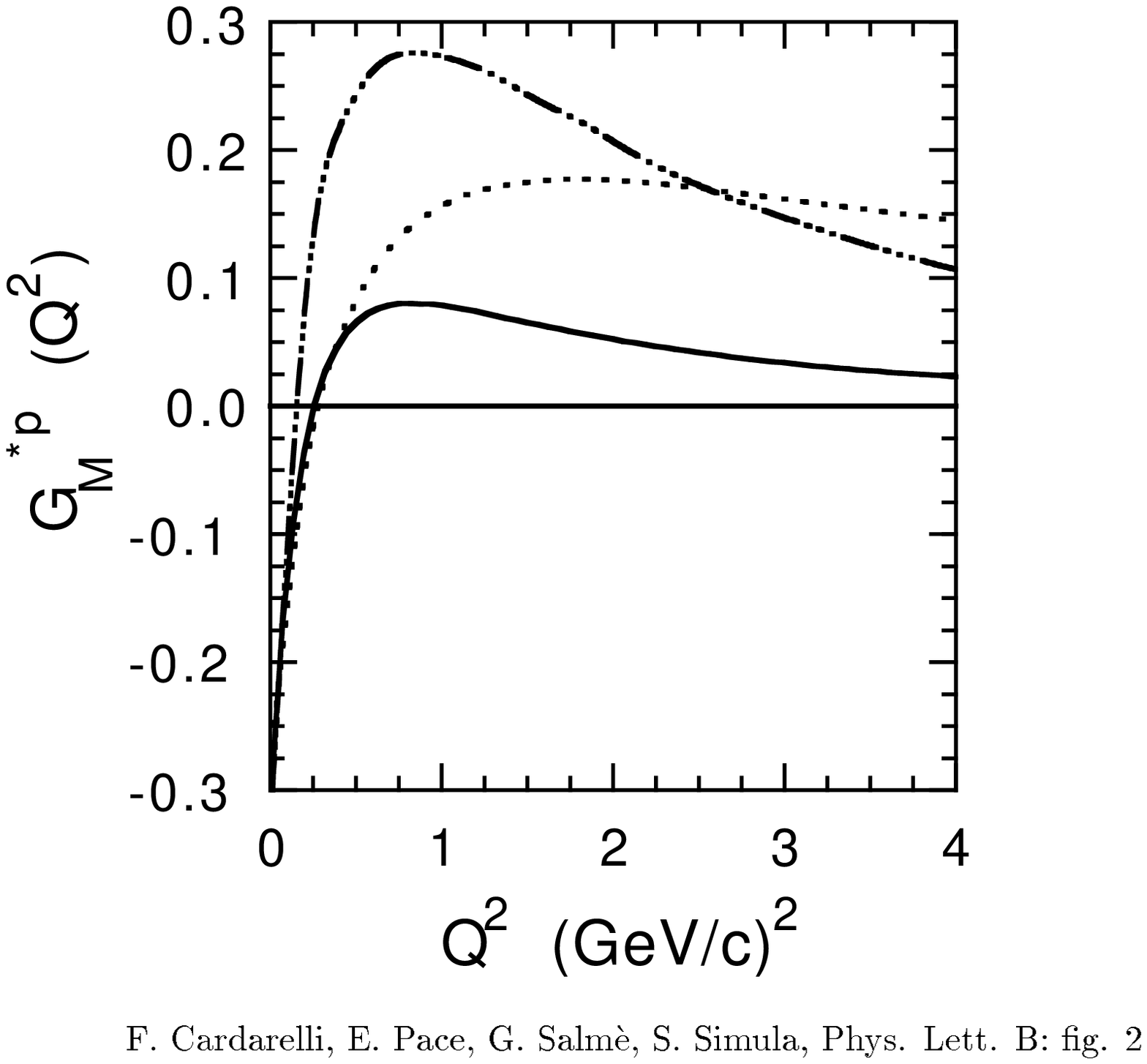}

\vspace{2.5cm}

\end{figure}

\newpage

\begin{figure}

\vspace{-4.0cm}

\epsfxsize=11.5cm \epsfig{file=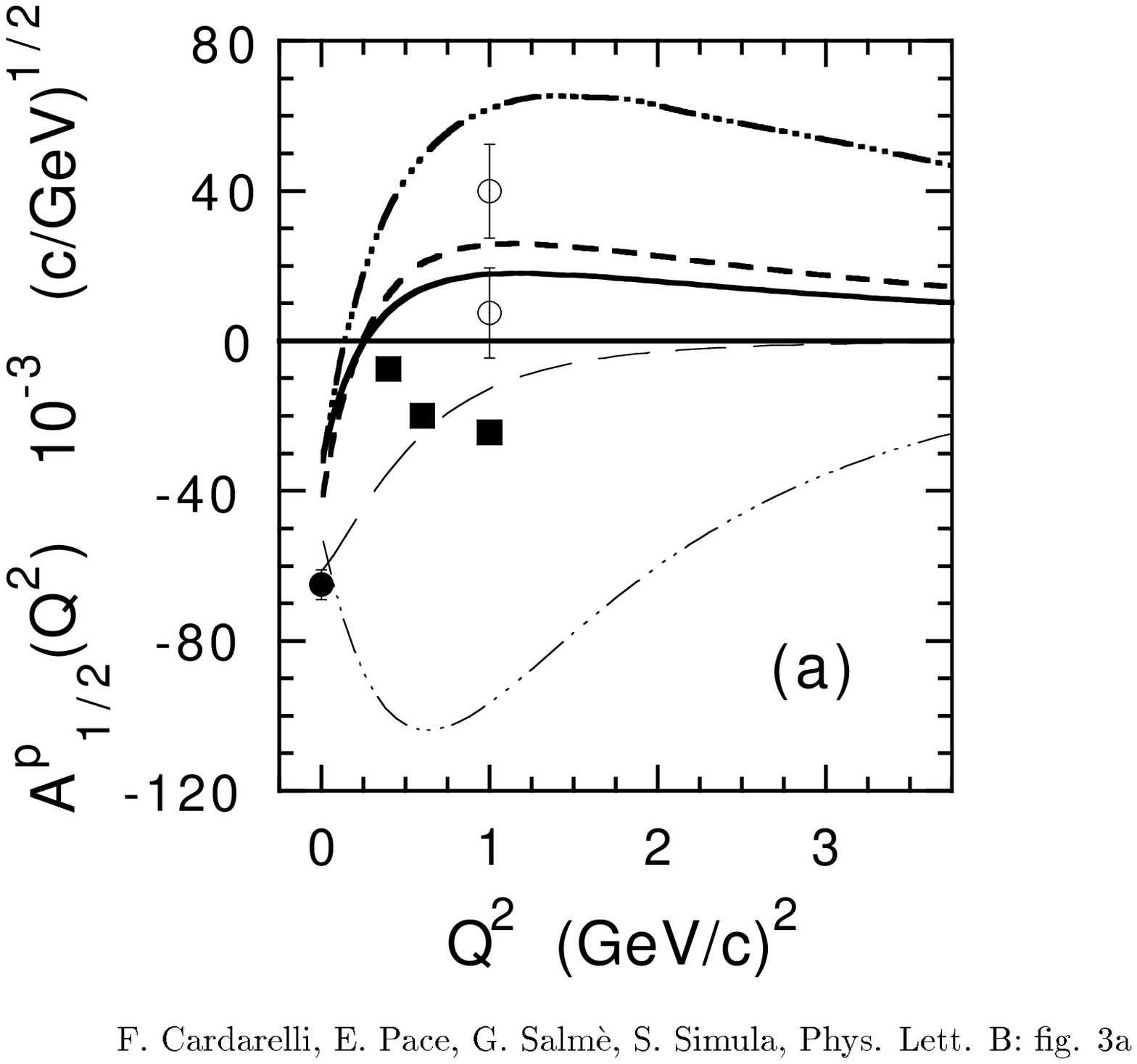}

\end{figure}

\begin{figure}

\vspace{-2.5cm}

\epsfxsize=11.5cm \epsfig{file=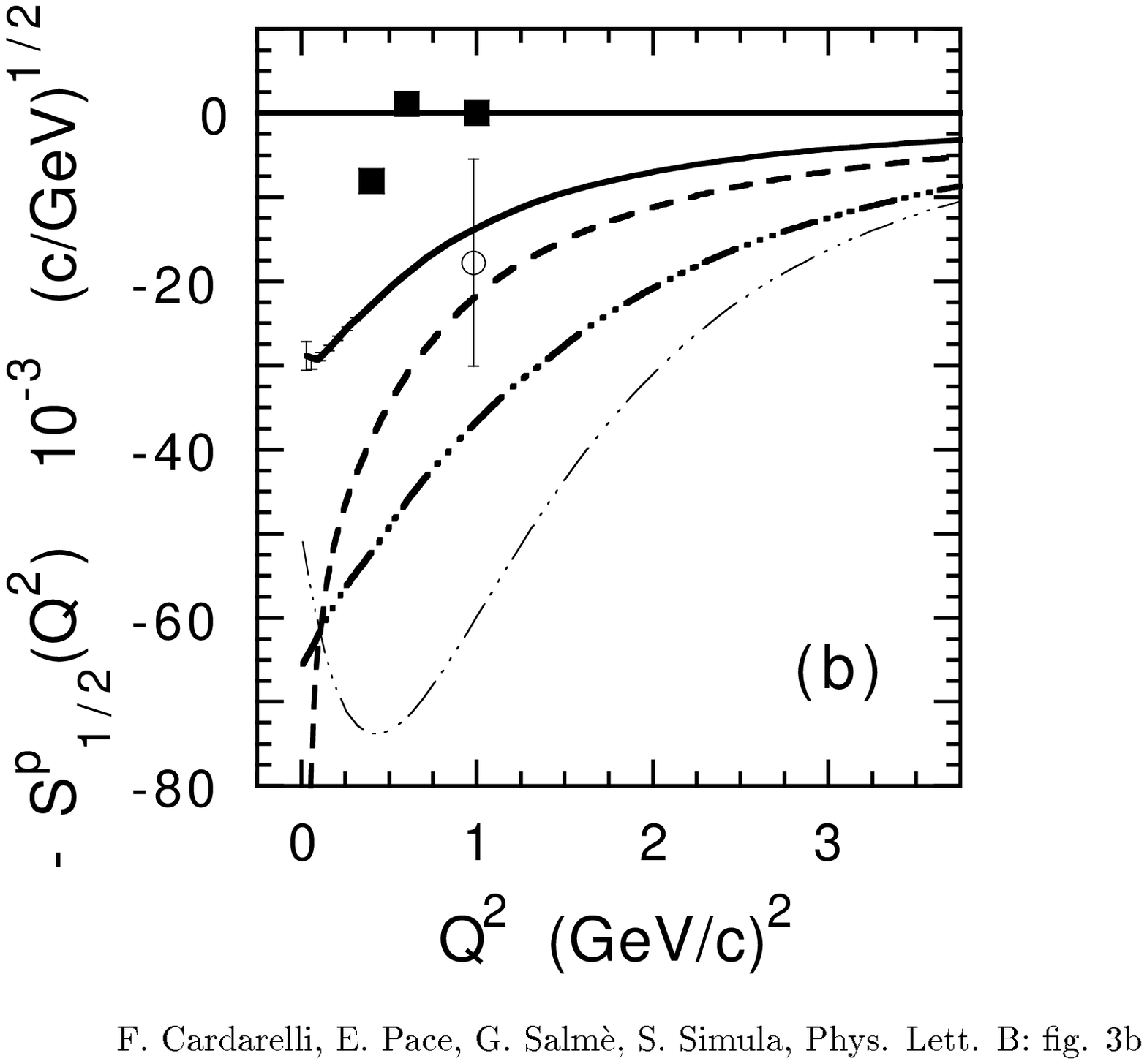}

\end{figure}

\newpage

\begin{figure}

\vspace{-4.0cm}

\epsfxsize=11.5cm \epsfig{file=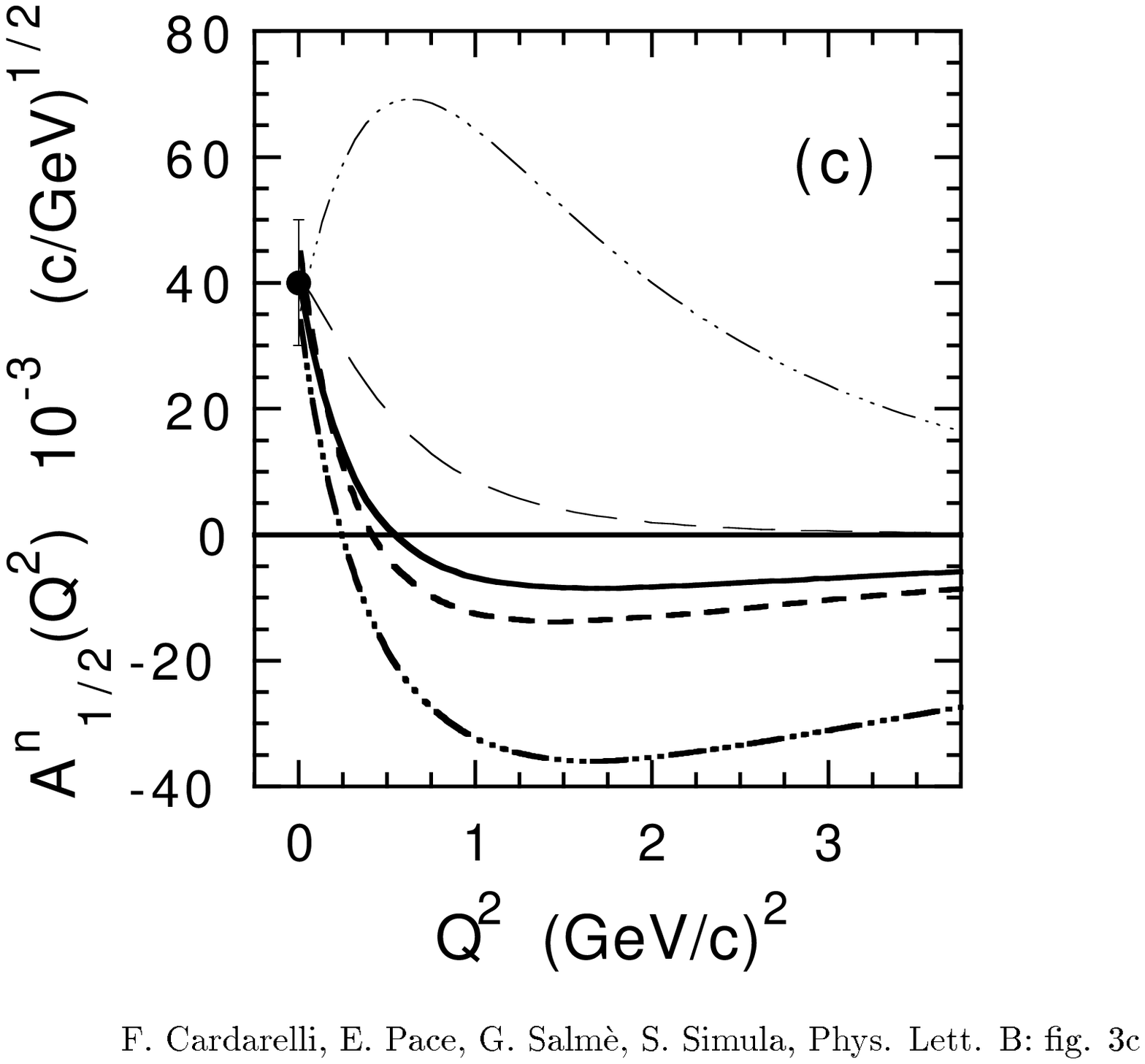}

\end{figure}

\begin{figure}

\vspace{-2.5cm}

\epsfxsize=11.5cm \epsfig{file=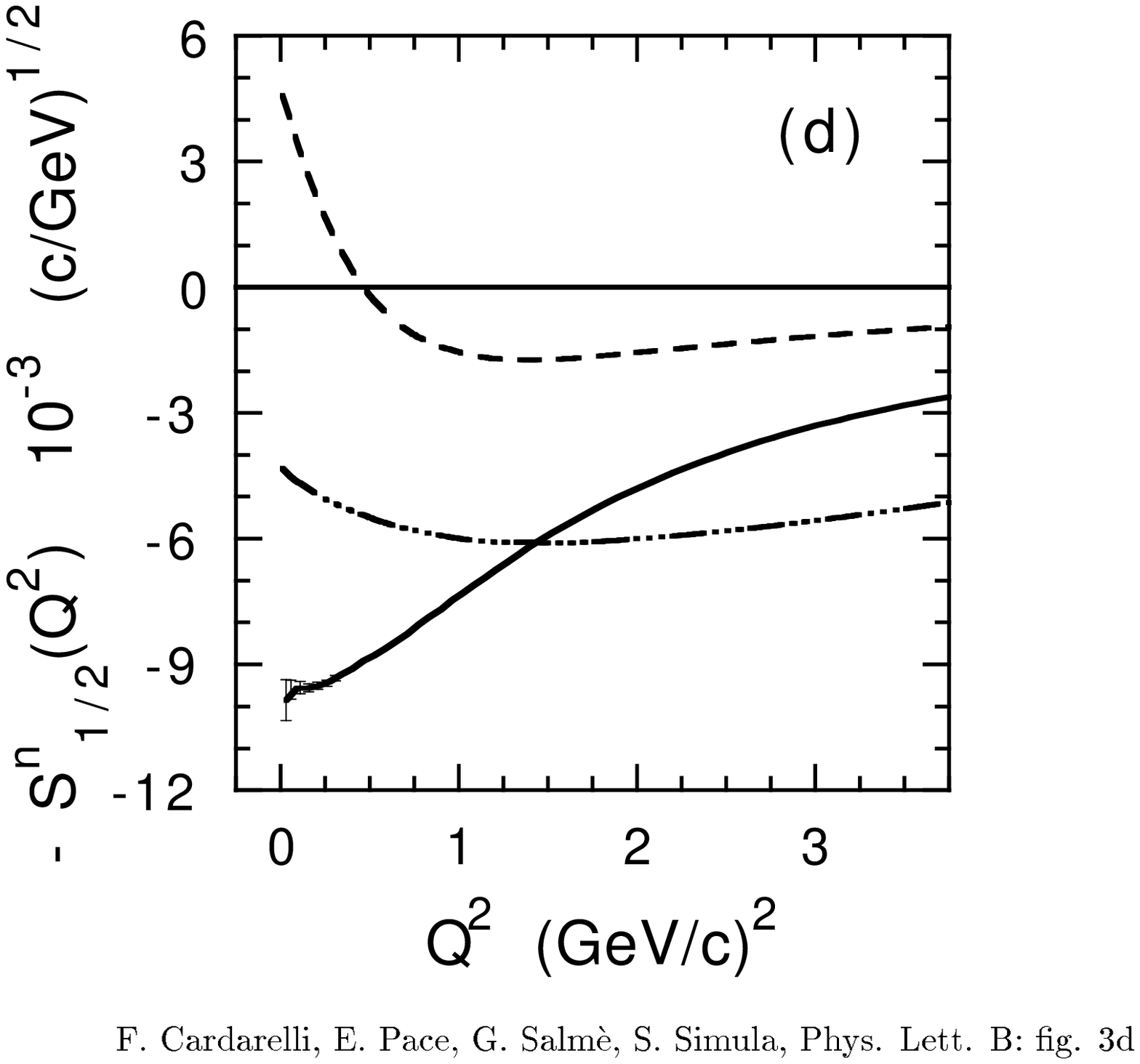}

\end{figure}

\newpage

\begin{figure}

\vspace{-5.0cm}

\epsfig{file=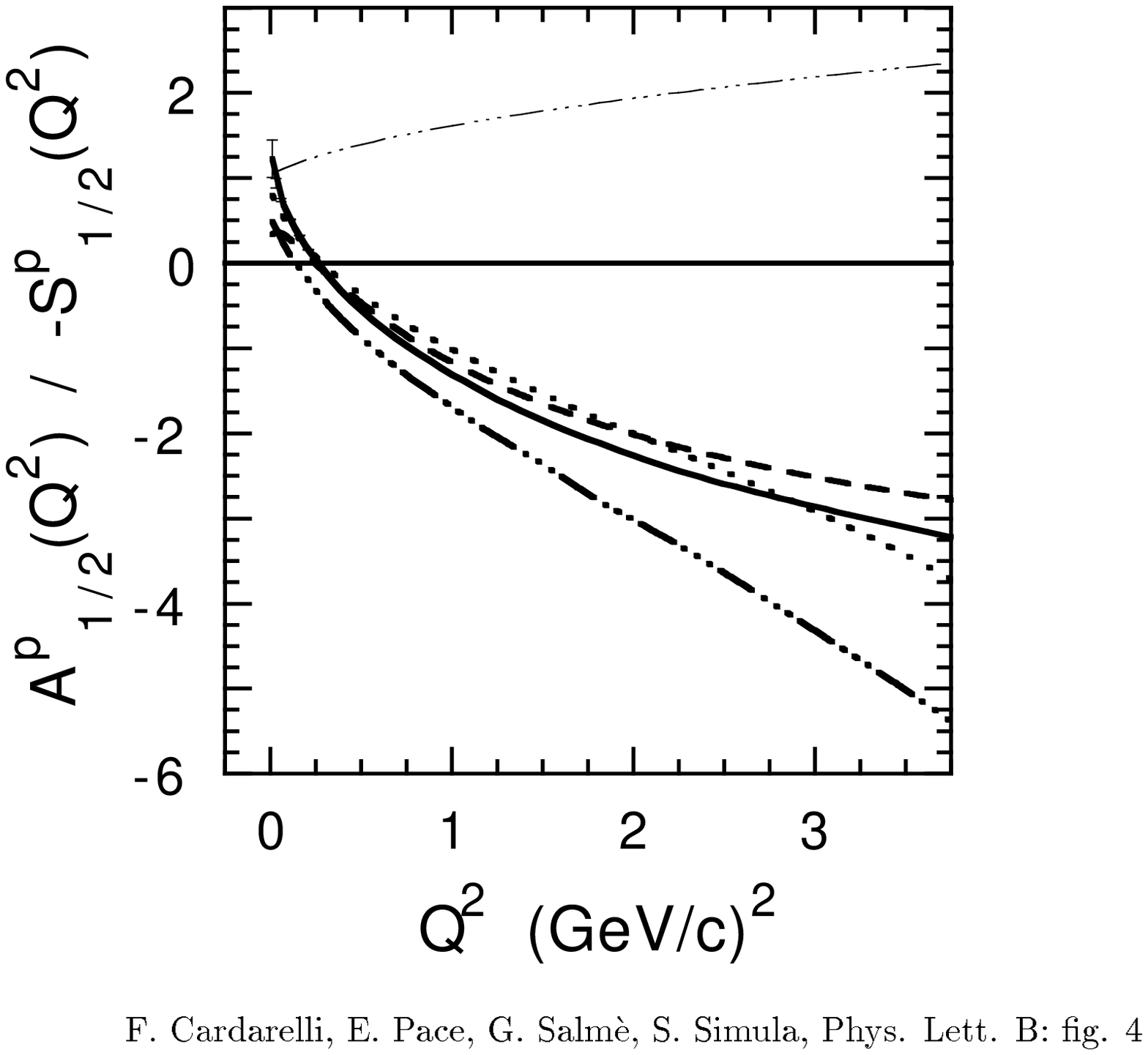}

\end{figure}

\end{document}